# Influence of hydrogen on paramagnetic defects induced by UV laser exposure in natural silica

**Fabrizio Messina, Marco Cannas**[*] and **Roberto Boscaino**

[1] Dipartimento di Scienze Fisiche ed Astronomiche dell'Università di Palermo and Istituto Nazionale per la Fisica della Materia, via Archirafi 36, 90123 Palermo, Italy

Diffusion limited reactions of point defects were investigated in amorphous $SiO_2$ exposed to UV laser light. Electron spin resonance and in situ absorption measurements at room temperature evidenced the annealing of E' centers and the growth of H(II) centers both occurring in the post-irradiation stage and lasting a few hours. These transients are caused by reactions involving molecular hydrogen $H_2$, made available by dimerization of radiolytic $H^0$.

## 1 Introduction

Hydrogen, in atomic ($H^0$) and molecular ($H_2$) form, is mobile inside the silica matrix even at room temperature and because of its reactivity with various point defects it may influence the optical properties of silica-based materials [1]. From the application point of view, $H_2$ loading of $SiO_2$ plays a relevant role in the fabrication of optical fibers for ultraviolet (UV) transmission or Bragg gratings in fibers [2]. Despite the considerable technological interest, several aspects regarding the interplay between UV photo-generation of defects and their interaction with hydrogen remain unclear. In this respect, annealing experiments on irradiated silica provide an useful method to study the diffusion limited reactions of induced defects [1,3]. In the present paper we deal with the transient behavior of two paramagnetic defects induced in silica by UV laser irradiation. The first is the E' center, which consists in an unpaired electron (•) localized on a Si bonded with three O (≡Si•) [4] and also exhibits an optical absorption (OA) band at 5.8 eV [5]. The second is the H(II) center whose structure is a Ge atom bonded with two O and a H (=Ge•-H) and is observed in ESR spectrum as a doublet split by 11.8 mT [6]. Our experimental approach is based on the jointed use of electron spin resonance (ESR) and in situ OA spectroscopic techniques aiming to probe the reactions of mobile $H_2$ at the site of these defects.

## 2 Experimental methods

Natural silica (fused quartz) EQ906 samples, 5×5×1 $mm^3$, were employed in our experiments. As received, this material has an OH content of ~20 ppm by weight, as measured from the amplitude of the IR band at 3600 $cm^{-1}$. Moreover it exhibits an OA band at 5.1 eV and two correlated emissions at 3.1 and 4.2 eV associated with the twofold coordinated Ge (=Ge••) [7]. Ge impurity is present in the material in amount of ~1 ppm by weight due to natural contamination of the unfused quartz used in the manufacture.

[*] Corresponding author: e-mail: cannas@fisica.unipa.it, Phone: +39 091 6234 220, Fax: +39 091 6162 461

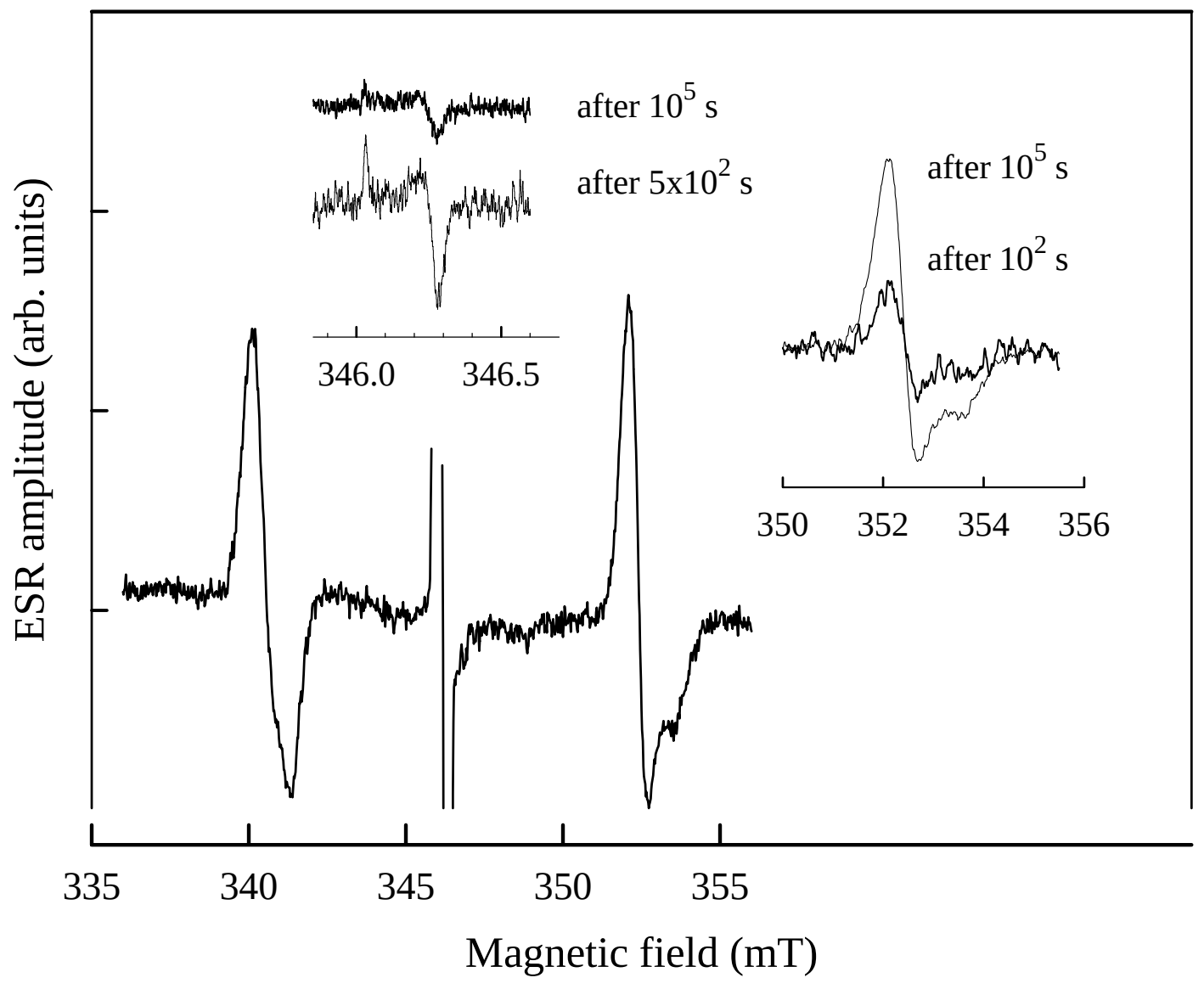

**Fig. 1** ESR spectra detected in the EQ906 sample irradiated by 2000 pulses showing the 11.8 mT doublet and the E' center signal recorded at different delays after the end of UV exposure.

Exposure with UV photons (hν≈4.7 eV) was performed at room temperature using the fourth harmonic generation from the pulsed radiation of a Nd :YAG laser (Quanta System SYL 201), at a repetition rate of 1 Hz, each pulse having energy density of W=40mJ/$cm^2$ and 5ns duration. UV-induced paramagnetic centers were investigated at room temperature by two techniques: in situ optical absorption (OA) and ex situ ESR spectroscopy. OA spectra were carried out by an optical fiber spectrophotometer (AVANTES S2000) equipped with a $D_2$ lamp source providing ~2 μW light power on the larger surface (5×5 $mm^2$) of the sample; the transmitted light was detected by a 2048 channels CCD linear array. Detection timing was triggered at the same repetition rate (1 Hz) of laser pulses; both during irradiation, when they were collected during each inter-pulse, and in the post-irradiation stage. ESR measurements were performed at different times (from $10^2$ s to $10^6$ s) after laser exposure on a spectrometer (Bruker EMX) working at 9.8GHz. Signals were detected with a 100 kHz modulation field, peak-to-peak amplitude 0.01 to 0.4 mT, while the microwave power was set low enough to prevent saturation. The concentration of revealed paramagnetic centers was evaluated by comparing the double-integrated ESR spectra with that of E' centers whose absolute density was determined with accuracy of ±20% by spin-echo measurements [8].

## 3 Results

UV induced ESR features detected in a sample at different delays after the exposure to 2000 laser pulses are shown in Fig. 1 and evidence the opposite behavior of the E' and H(II) centers. The amplitude of the E' center resonance line, in the central part of the spectrum, partially decreases by a factor ~2.5 on increasing the delay time up to $10^5$ s. As regards the H(II) centers, we observe a post-irradiation growth by looking at the high field component of the 11.8 mT doublet. Their concentration at the end of irradiation is estimated to be ~$6\times10^{14}$ $cm^{-3}$ and increases up to ~$2\times10^{15}$ $cm^{-3}$ after $10^5$ s. As pointed out in previous works [9], the growth of H(II) centers is correlated with the bleaching of the absorption at 5.1 eV and emissions at 3.1 and 4.2 eV associated with the twofold coordinated Ge.

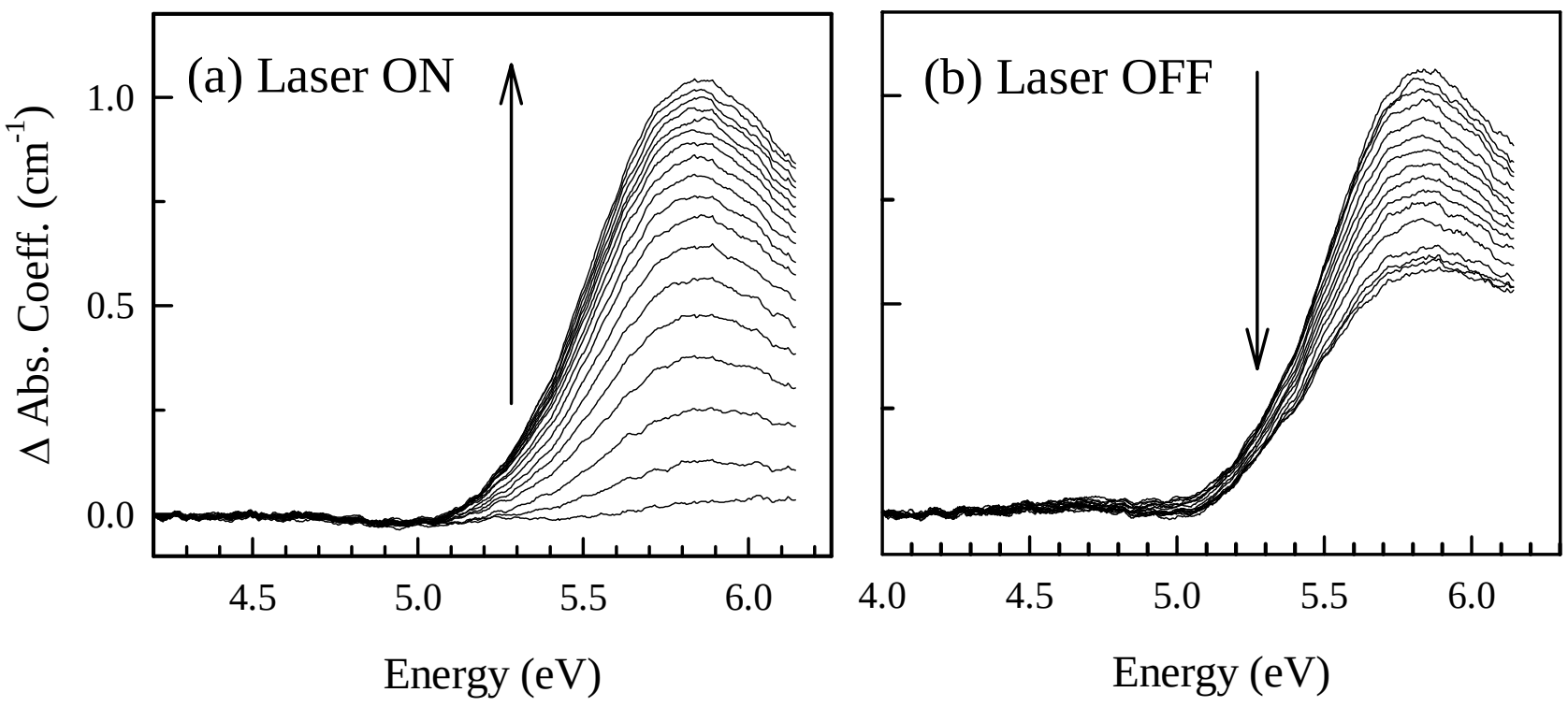

**Fig. 2** Difference OA spectra detected in situ during 2000 s laser exposure (a) and during 3600 s in the post-irradiation stage (b). Arrows indicate the evolution of OA spectra with increasing time.

The generation and the post-irradiation decay of E’ centers are also observed by in situ OA measurements. Fig. 2 shows the difference OA spectra detected in another sample both during irradiation with 2000 laser pulses and in the post-irradiation stage. UV laser photons induce the band centered at 5.83±0.03 eV associated with the E’ centers. At the end of irradiation the 5.8 eV band amplitude is 1.08±0.02 $cm^{-1}$ which corresponds to a concentration of $1.7\times10^{16}$ $cm^{-3}$ on the basis of its cross section value ($\sigma$=$6.4\times10^{-17}$ $cm^2$) [10]. When the laser is switched off, the 5.8 eV band partially decreases with time, the reduction being ~40% in the first hour since the end of irradiation. This is consistent with the post-irradiation reduction of the E’ center ESR signal.

Fig. 3 summarizes for the two defects the absolute concentration variation in the post-irradiation stage from the value measured at the end of irradiation. We observe that the kinetics of both defects is almost completed within a few hours; the concentration of annealed E’ centers is ~$10^{16}$ $cm^{-3}$, the growth of H(II) centers is ~$1.4\times10^{15}$ $cm^{-3}$.

## 4 Discussion and conclusions

The above reported results evidence that two kinds of paramagnetic defects are induced by UV radiation: E’ and H(II) centers. The generation of E’ centers is commonly associated with the bond cleavage at a site of precursor defects such as oxygen deficient centers (ODC), strained Si-O bonds, or impurity bonds (Si-H, Si-Cl) [2]. Present results do not allow to clarify unambiguously the generation process. We just note that 4.7 eV laser radiation really excites OA bands related to ODC, whereas the electronic transitions related to the other precursors have not been yet clarified and a multi-photon absorption of laser radiation could be involved in our experiments. A fraction of E’ center decays after laser exposure with a transient lasting ~$10^4$ s at room temperature. This timescale is consistent with the diffusion parameters of molecular hydrogen inside $SiO_2$. In fact, taking into account the diffusion coefficient of $H_2$ at T=300 K (D=$2.3\times10^{-11}$ $cm^2$ $s^{-1}$) [11], within t~$10^4$ s the root mean square of diffusion length $(\pi Dt)^{1/2}$ increases up to $10^{-3}$ cm. Hence, comparing with the maximum concentration of E’ centers found at the end of irradiation (C≈$1.7\times10^{16}$ $cm^{-3}$), we get that $H_2$ moves up to ~$10^2$ times the average distance $C^{-1/3}$ between two E’ centers. Then, the observed isothermal annealing of E’ centers can be explained by the following process:

$$\equiv Si^{\bullet} + H_2 \rightarrow \equiv Si\text{-}H + H^0 \quad (1)$$

where the released $H^0$ can be involved in further reactions with E’ centers or other defects.

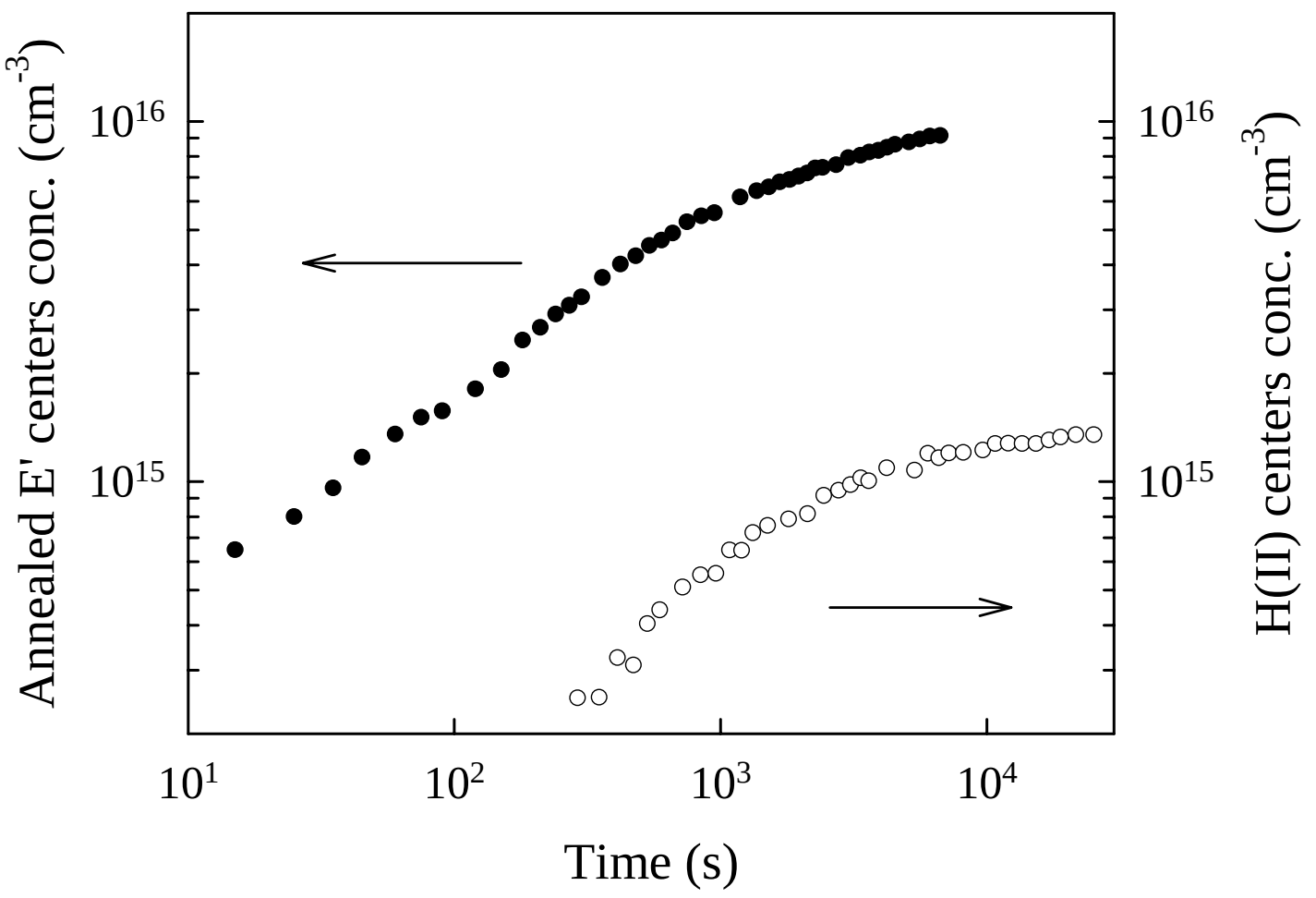

**Fig. 3** Variation of concentration of E' centers (full symbol) and H(II) centers (empty symbol) in the post-irradiation stage starting from 10 s after the end of irradiation.

The hydrogen involved in reaction (1) is produced from OH and SiH bonds whose UV-induced breaking makes free $H^0$ which, at room temperature, diffuses very quickly and dimerizes to form $H_2$ [1, 3].

As regards the growth of H(II) centers, it has been associated with the conversion of twofold coordinated Ge through a two-step process [12]. The first step is the cracking of $H_2$ at the site of a paramagnetic center which releases a $H^0$ atom; in the present experiment this corresponds to reaction (1). The second step is the trapping of $H^0$ at the site of a twofold coordinated Ge:

$$=\mathrm{Ge}^{\bullet\bullet} + H^0 \rightarrow =\mathrm{Ge}^{\bullet}\text{-H} \qquad (2)$$

Hence, the growth kinetics of H(II) centers is mainly governed by reaction (1) and occurs within the same timescale, as evidenced by the results of Fig. 3.

In conclusion, the reaction of diffusing $H_2$, of radiolytic origin, with UV-induced E' centers causes their annealing in the post-irradiation stage. The reaction makes available $H^0$ atoms which can be trapped at the site of twofold coordinated Ge so producing H(II) centers.

**Acknowledgements** The authors thank S. Agnello, G. Buscarino, F. M. Gelardi for useful discussions and G. Lapis and G. Napoli for technical assistance. This work is part of a national project (PRIN2002) supported by the Italian Ministry of University Research and Technology.